\documentclass
[aps,prl,twocolumn,superscriptaddress,showpacs,floatfix]{revtex4-1}%
\usepackage{graphicx,amsmath,amssymb}
\usepackage{amsmath}
\usepackage{amsfonts}
\usepackage{amssymb}
\usepackage{graphicx}%
\providecommand{\U}[1]{\protect\rule{.1in}{.1in}}
\begin{document}
\title{Non-equilibrium fractional quantum Hall states visualized by optically
detected MRI}
\author{John N. Moore}
\affiliation{Department of Physics, Tohoku University, Sendai 980-8578, Japan}
\author{Junichiro Hayakawa}
\affiliation{Department of Physics, Tohoku University, Sendai 980-8578, Japan}
\author{Takaaki Mano}
\affiliation{National Institute for Materials Science, Tsukuba, Ibaraki 305-0047, Japan}
\author{Takeshi Noda}
\affiliation{National Institute for Materials Science, Tsukuba, Ibaraki 305-0047, Japan}
\author{Go Yusa}
\email{yusa@m.tohoku.ac.jp}
\affiliation{Department of Physics, Tohoku University, Sendai 980-8578, Japan}
\date{\today
}

\begin{abstract}
Using photoluminescence microscopy enhanced by MRI, we visualize in real space
both electron and nuclear polarization occurring in non-equilibrium FQH
liquids. We observe stripe-like regions comprising FQH excited states which
discretely form when the FQH liquid is excited by a source-drain current.
These regions are topologically protected and deformable, and give rise to
bidirectionally polarized nuclear spins as spin-resolved electrons flow across
their boundaries.

\end{abstract}

\maketitle

The~study of nuclear spins in semiconductor materials has become increasingly
relevant as both a tool for probing the properties of electrons in
semiconductors
\cite{tycko,barrett,gammon,kuzma,kikkawa,salis,kumada,freytag,smet,tiemann,kronmuller,kronmuller99}%
, and a means for developing understanding of quantum information processing
\cite{kane,vandersypen,yusa,petta,koppens,childress}. These studies were
enabled by unconventional nuclear magnetic resonance (NMR) in which nuclear
polarization is detected via resistance or optical response of samples, i.e.
resistively or optically detected NMR.

In particular, resisitively detected NMR has been a powerful tool for studying
fractional quantum Hall (FQH) states
\cite{kronmuller,kronmuller99,freytag,smet,kumada,tiemann}. FQH states form in
a strongly interacting 2D electron system under a perpendicular magnetic field
at fractional values of the Landau level filling factor $\nu$ \cite{tsui}. The
FQH state is a type of topological state of matter, and owes its energetic
stability to its boundaries, i.e. edge states, which exist between the bulk
state and the surrounding vacuum \cite{wen}. In the trivial case, e.g.,
$\nu=1/3$, the 2D bulk of FQH states is gapped by electron-spin- (Zeeman
energy) and Coulomb-interaction-induced energy gaps \cite{tsui}. However,
there also are non-trivial FQH states, e.g. $\nu=2/3$, for which the gap can
be closed at a certain critical magnetic field $B_{\text{C}}$ under which the
non-magnetic (unpolarized) and perfect ferromagnetic (polarized) phases,
corresponding to electron spin polarizations $P$ of $0$ and $1$, respectively,
are degenerate; this leads to a first-order phase transition and the formation
of magnetic domains \cite{hayakawa}. For $\nu$ slightly lower (higher) than
$2/3$, the ferromagnetic (non-magnetic) phase becomes the ground state.

Near the phase transition, it is known that longitudinal resistance $R_{xx}$
is dramatically increased in response to a large source-drain current
\cite{kumada,freytag,smet,tiemann,kronmuller,kronmuller99,yusa} bringing the
system out of equilibrium. This $R_{xx}$ enhancement is understood in
conjunction with the nuclear spin polarization $P_{\text{N}}$, as confirmed by
resistively detected NMR\cite{kronmuller99}. However, the underlying mechanism
behind the $R_{xx}$ enhancement and the associated $P_{\text{N}}$ has been
unknown for decades despite its importance and versatility. In this Letter, we
examine the origins of this intriguing effect in relation to the topological
stability of the non-equilibrium FQH state. Using photoluminescence (PL)
microscopy enhanced by magnetic resonance imaging (MRI), we visualize in real
space both electron and nuclear polarization occurring in non-equilibrium FQH
liquids. In response to a perturbation, we show the discrete transformation of
the topologically protected FQH liquid into stripe-like regions comprising FQH
excited states. These regions constitute topologically protected structures
across which spin-resolved electrons flow to generate bidirectionally
polarized nuclear spins. These findings demonstrate powerful and sensitive
microscopic visualization tools for exploring spin-related phenomena and
furthering understanding of semiconductor physics.

\begin{figure*}[t]
\par
\begin{center}
\includegraphics{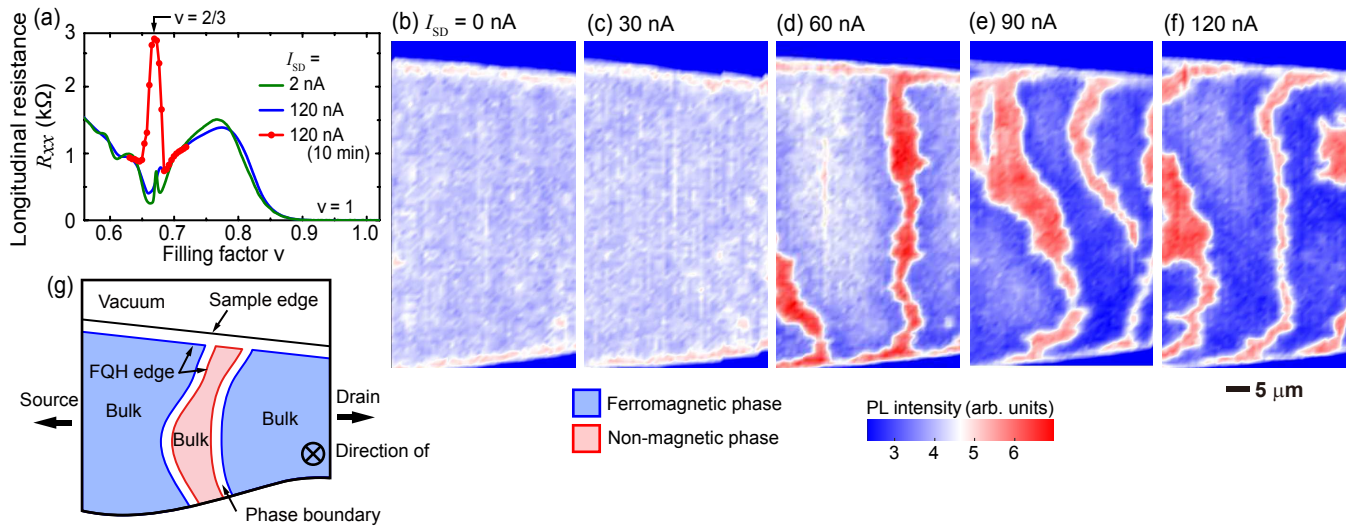}
\end{center}
\caption{(Color online) (a) Longitudinal resistance $R_{xx}$ as function of
$\nu$. (Green) $I_{\text{SD}}=2$, (blue, red) $120$~nA. $\nu$ is scanned over
$0.5-1.02$ with $3.15\times10^{-4}$ /s (green, blue) and $6.18\times10^{-6}$
/s (red) scan speed. (b) to (f) $38\times$ $68$-$\mu$m$^{2}$ spatial images
showing PL intensity of charged exciton singlet peak (PL intensity), at $\nu=$
$0.660$ with $I_{\text{SD}}$ (Hall voltage, $V_{\text{H}}=\frac{3}{2G_{0}%
}I_{\text{SD}}$, $G_{0}$: quantized conductance) of: (b) $0$ ($0$); (c) $30$
($1.16$); (d) $60$ ($2.32$); (e) $90$ ($3.48$); and (f) $120$ nA ($4.65$ meV).
$T\sim60$ mK and $B=B_{\text{C}}=$ $6.8$ T throughout unless otherwise
specified \cite{SI}. (g) Schematic of observed images. The microscopic
internal structure of the $2/3$ edge state is not shown here, because it
remains unclear \cite{bid,wu}.}%
\label{fig:fig1}%
\end{figure*}

First, we confirmed the previously reported $R_{xx}$ enhancement using a
$15$-nm-wide GaAs/AlGaAs quantum well (QW) sample containing a FQH liquid
\cite{SI}. Under the small perturbation of an alternating source-drain current
$I_{\text{SD}}$ ($2$~nA, $13$~Hz) at approximately $B_{\text{C}}$, $\nu=2/3$,
$R_{xx}$ becomes low [Fig.~1(a), green line] with a small peak due to the
phase transition\cite{kronmuller,kronmuller99}. In contrast, under large
perturbation ($I_{\text{SD}}$ = $120$~nA), $R_{xx}$ is dramatically increased
in the steady state [Fig.~1(a), red line], similar to previous reports
\cite{kumada,freytag,smet,tiemann,kronmuller,kronmuller99,yusa}.

Under these non-equilibrium conditions we obtained real-space images of FQH
liquids [Figs.~1(b)$-$1(f)] under varied $I_{\text{SD}}$-induced perturbations
at $B=B_{\text{C}}=6.8$~T via polarization-sensitive scanning optical
microscopy and spectroscopy of the sample's PL \cite{hayakawa,SI}. The PL
intensity from the singlet charged excitons \cite{yusaPRL,wojs} used to
construct these images is primarily anti-correlated with the local $P$
\cite{hayakawa}. Therefore, non-magnetic and ferromagnetic phases are
distinguished by strong and weak PL intensities, respectively \cite{hayakawa}.
Because the ground state at $\nu=0.660$ ($<2/3$) is a ferromagnetic phase, the
PL of the entire region is uniformly low (blue) for low perturbation
[$I_{\text{SD}}\leq30$~nA; Figs.~1(b) and 1(c)]. Gapless edge channels at the
bulk-vacuum interface protect the gapped bulk FQH liquid, although its
electrochemical potential is far from equilibrium owing to the Hall voltage,
$V_{\text{H}}$. For high perturbation, high-PL-intensity regions (red) appear
(determined to indicate a non-magnetic phase based on their spectra
\cite{hayakawa}) and bridge both sides of the sample. Importantly, these
stripe-shaped structures comprising the excited FQH liquid extend over a
macroscopic scale and are topologically protected at their boundaries; because
the boundary of two coexisting phases requires spin flips, the exchange
interaction induces an energy barrier at the boundary \cite{shibata},
producing a new type of edge [Fig.~1(g)]. Through these
exchange-interaction-induced edges, electrons injected from an electrode can
return to the electrode (backscattering along the boundary), contributing to a
non-zero $R_{xx}$ [Fig.~1(a)].

This backscattering creates a local electrochemical potential difference
across the two phases, generating potential steps at each phase boundary that
descend monotonically in the direction opposing the conventional current. The
reasonably well-reproducible stripe patterns \cite{SI} indicate that these
potential steps are comparable to fluctuations in the sample's intrinsic
random potential ($\sim100~\mu$eV \cite{hayakawa,SI}). This, in turn, suggests
that the backscattering rate at a boundary is $<\sim0.1$ using
Landauer-B\"{u}ttiker formalism \cite{datta}, even without considering the
internal structures of the ferromagnetic- and non-magnetic-phase $2/3$ edge
states, which remain unclear \cite{bid,wu}. Note that there are spin-resolved
electrons flowing along the edges, and almost all the electrons must cross
phase boundaries to contribute to $I_{\text{SD}}$ (forward scattering across
the boundary), with half undergoing spin flipping.

\begin{figure*}[t]
\par
\begin{center}
\includegraphics{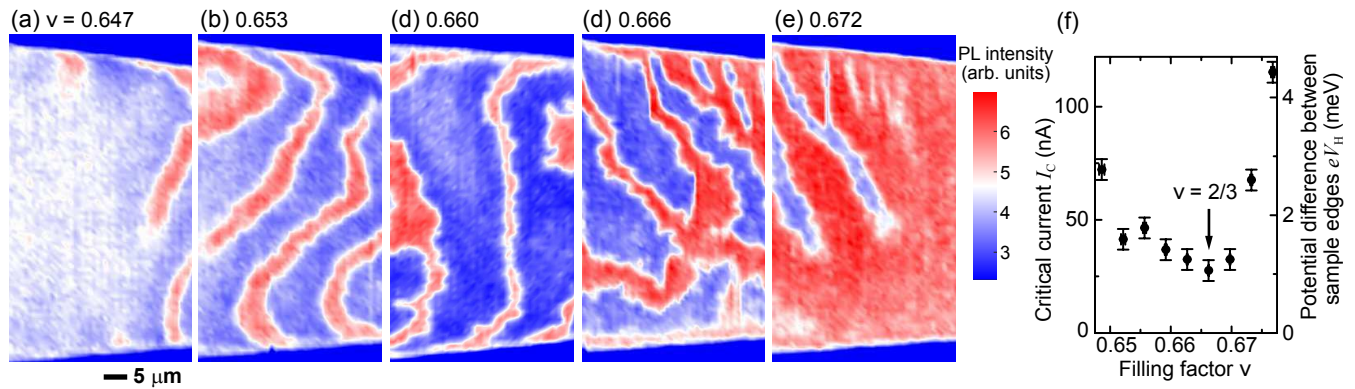}
\end{center}
\caption{(Color online) (a) to (e) $38\times$ $68$-$\mu$m$^{2}$ PL-intensity
spatial images at $\nu$ of: (a) $0.647$; (b) $0.653$; (c) $0.660$; (d)
$0.666$; and (e) $0.672$, with $I_{\text{SD}}=120$ nA. (f) Critical current at
which first stripe becomes visible ($I_{\text{C}}$) as function of $\nu$.
Error in $I_{\text{C}}$ determined by step size of applied current. Error in
$\nu$ calculated from uncertainty in electron density. Right $y$-axis:
Potential difference between sample edges, $eV_{\text{H}}$.}%
\label{fig:fig2}%
\end{figure*}

Such structures can be observed over a relatively wide range of $\nu
=2/3\pm\sim3\%$ [Figs.~2(a)$-$2(e)], unlike the phase transition ($\nu
=2/3\pm\sim0.3\%$\cite{hayakawa}). At $\nu=0.666$, the areas of the two phases
are almost equal, owing to the phase transition [Fig.~2(d)], and the minimum
current (critical current; $I_{\text{C}}$) required to excite the stripe
patterns is at a minimum [Fig.~2(f)]. For $\nu<(>)$ $2/3$ , the stripes are in
the excited, i.e., non-magnetic (ferromagnetic) phase [Figs.~2(a)$-$2(c) and
2(e)]. On either side of $\nu=2/3$, $I_{\text{C}}$ increases sharply; thus,
$I_{\text{C}}$ corresponds to the minimum energy required to flip a large
amount of electron spins simultaneously to form the energetically unfavorable
long-range-ordered structures.

\begin{figure}[b]
\par
\begin{center}
\includegraphics{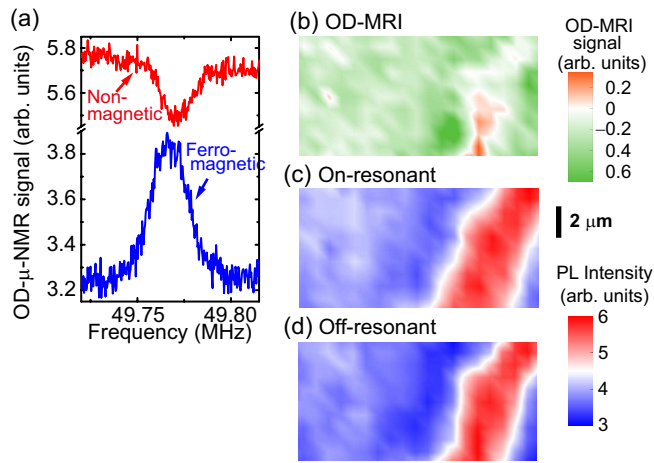}
\end{center}
\caption{(Color online) (a) Optically detected (OD)-$\mu$-NMR spectra. (b)
$15\times7.5$-$\mu$m$^{2}$ OD-MRI. PL-intensity spatial images under (c)
on-resonant ($49.7717$ MHz) and (d) off-resonant r.f. ($49.84$ MHz). Data were
obtained at $\nu=0.660$ and $T=$ $\sim80$ mK. The r.f. power at the generator
for both methods was $-3.8$ dBm.}%
\label{fig:fig3}%
\end{figure}

$R_{xx}$ enhancement caused by backscattering is known to be linked to nuclear
spin polarization as mentioned above
\cite{freytag,smet,tiemann,kronmuller99,yusa}. Thus, we performed optically
detected microscopic NMR (OD-$\mu$-NMR) to examine the relation between these
stripe structures and $P_{\text{N}}$. The $\mu$-PL intensity was measured
while irradiating the sample with continuous wave r.f. from a two-turn coil
wrapped around the sample. The OD-$\mu$-NMR [Fig.~3(a)] is the $\mu$-PL
intensity plotted as a function of the irradiating r.f. frequency. This
spectrum measured in the ferromagnetic (non-magnetic) phase shows a clear
resonant peak (dip) corresponding to $^{75}$As nuclei [Fig.~3(a)]. The $\sim
5$-kHz frequency difference between the peak and dip is caused by the
difference in $P$ between $1$ and $0$, respectively, which is consistent with
the Knight shift obtained via resistively detected NMR \cite{yusaCondMat}. The
optically detected MRI (OD-MRI) image [Fig.~3(b)] shows the PL-intensity map
corresponding to that collected during the \textit{on}-resonant r.f.
[Fig.~3(c)] minus that collected during the \textit{off}-resonant r.f.
[Fig.~3(d)]. In other words, this image displays the spatial distribution in
the sizes of resonant PL peak (green regions) and dip (orange region)
occurring in the ferromagnetic and non-magnetic phases, respectively [Fig.~3(b)].

The PL images around $\nu=2/3$ contain contributions from the spatial
distributions of both $P$ and $P_{\text{N}}$. To extract the contribution of
$P_{\text{N}}$ alone, we performed pump-probe imaging at $\nu=0.5$, where $P$
is homogeneously $\sim0.5$ \cite{tiemann}. The pump-probe images shown in
Figs.~4(a) and 4(b), were obtained via the following procedure: (i) The $\mu
$-PL collection spot was moved to an initial point; (ii) The pump condition
was set ($\nu=0.660$ for $4$~min with $I_{\text{SD}}$ $=120$~nA); (iii) The
$\mu$-PL spectrum was obtained; (iv) The probe condition ($\nu=1/2$ with
$I_{\text{SD}}$ $=0$~nA) was set; (v) After $3$~s, the $\mu$-PL spectrum was
obtained; and (vi) The collection spot was moved to the next point. We
repeated this procedure while raster scanning the sample to construct the
images under the pump [Fig.~4(a)] and probe [Fig.~4(b)] conditions. A stripe
structure, which is generated by pumping the current at $I_{\text{SD}}=120$~nA
and $\nu=0.660$ [Fig.~4(a)], remains visible $3$~s after current deactivation
($I_{\text{SD}}=0$) and changing to $\nu=0.5$ [Fig.~4(b)]. Under this
condition, $P$ does not contribute significantly to the spatial pattern.
Rather, the pattern seen here results from the remaining $P_{\text{N}}$. A
sufficient pause with no current pumping causes the pattern to disappear
[Fig.~4(c)]. The longitudinal relaxation time $T_{1}$ of the nuclear spins at
$\nu=0.5$ is estimated to be on the order of a minute, based on the PL time
dependence measured at two fixed points in the two phases [Fig.~4(d)].

Both OD-MRI and pump-probe imaging clearly show that the nuclear spins are
polarized on both sides of the phase boundaries, but with opposite polarities.
This asymmetry across the boundary is expected for forward-scattering
electrons contributing to $I_{\text{SD}}$ which flip their spins after
crossing into an adjacent domain. We conclude, therefore, that $P_{\text{N}}$
is parallel (anti-parallel) to $B$ in the non-magnetic (ferromagnetic) phase.
These images also reveal that $P_{\text{N}}$ within $\sim5~\mu$m of the
boundary tends to be slightly, yet noticeably, stronger than that far from the
boundary [Figs.~3(b) and 4(b)], a tendency clearly visible in many of the
$\nu\sim2/3$ PL images having stripes. We reason that this behavior is due to
$P_{\text{N}}$ diffusion with a typical diffusion length of $\sim5-10~\mu$m.
We cite the spatial distribution of $P_{\text{N}}$ as likely having a
stabilizing effect on the stripe shapes over time. The generation of
$P_{\text{N}}$ also modifies the electrochemical potential landscape via the
Overhauser effect, which in turn modifies the rates of backscattering along
the stripe boundaries.

\begin{figure}[t]
\par
\begin{center}
\includegraphics{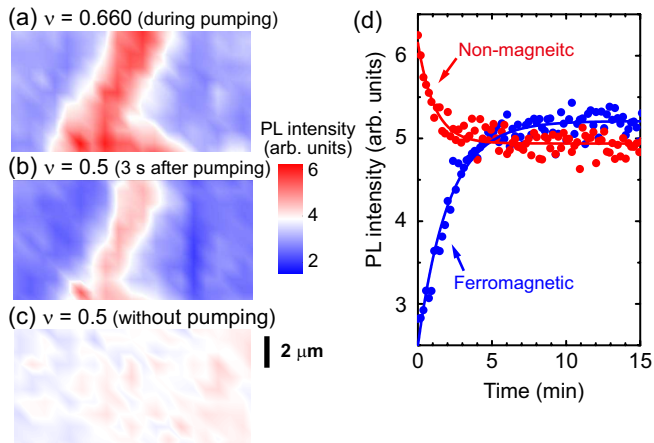}
\end{center}
\caption{(Color online) Pump-probe images and $\mu$-PL intensity time
evolution. (a) to (c) $15\times7.5$-$\mu$m$^{2}$-PL intensity spatial images
at; (a) $\nu=0.660$ during pumping, $I_{\text{SD}}=120$ nA; (b) $\nu=0.5$ $3$
s after pumping; and (c) $\nu=0.5$, $I_{\text{SD}}=0$ nA. (d) PL intensity
from two fixed points at (red) non-magnetic and (blue) ferromagnetic phases.
$t=0$: time when $\nu$, $I_{\text{SD}}$ were adjusted from pumping ($0.660$,
$120$ nA) to probing ($0.5$, $0$ nA) conditions. Each data point in
Fig.~$4$(d) was fitted using $a\exp\frac{t}{T_{1}}+b$, where $a=-2713.8$
($1216.5$), $T_{1}=2.059$ ($1.127$) min, and $b=5209.6$ ($4938.8$) for the
ferromagnetic (non-magnetic) phase.}%
\end{figure}

The findings presented here give insight into the qualities of dynamic nuclear
polarization (DNP) generated in non-trivial FQH liquids. The important role of
topological protection in non-equilibrium phenomena of these liquids is also
illuminated. At the same time, this study offers a form of MRI which surpasses
conventional techniques in its sensitivity to polarization magnitude and
direction and its spatial resolution. We believe these findings and methods
are directly relevant to efforts in quantum engineering based on DNP, and to
topological materials, extending to fractional topological insulators
\cite{levin} and topological quantum computation \cite{kitaev}.

\begin{acknowledgments}
The authors are grateful to N. Shibata, K. Muraki, K. Nomura, N. Kumada, and
T. Fujisawa for discussions, and to Y. Hirayama and M. Matsuura for
experimental support. This work was supported by the Mitsubishi Foundation,
and a Grant-in-Aid for Scientific Research (no. 24241039) from the Ministry of
Education, Culture, Sports, Science, and Technology (MEXT), Japan. J.N.M. was
supported by a Grant-in-Aid from MEXT and the Marubun Research Promotion
Foundation. J. H. was supported by a Grant-in-Aid from the Tohoku University
International Advanced Research and Education Organization.
\end{acknowledgments}

\end{document}